\documentclass[aps,prd,superscriptaddress,preprint,tightenlines,nofootinbib,floatfix]{revtex4}

\usepackage{graphicx} 
\usepackage{dcolumn}  
\usepackage{bm}       

\begin{document}

\preprint{CLNS 10/2069}  
\preprint{CLEO 10-06}    

\title{\boldmath Analysis of the Decay $D^0 \rightarrow K^0_S\pi^0\pi^0$}

\author{N.~Lowrey}
\author{S.~Mehrabyan}
\author{M.~Selen}
\author{J.~Wiss}
\affiliation{University of Illinois, Urbana-Champaign, Illinois 61801, USA}
\author{J.~Libby}
\affiliation{Indian Institute of Technology Madras, Chennai, Tamil Nadu 600036, India}
\author{M.~Kornicer}
\author{R.~E.~Mitchell}
\author{M.~R.~Shepherd}
\author{C.~M.~Tarbert}
\affiliation{Indiana University, Bloomington, Indiana 47405, USA }
\author{D.~Besson}
\affiliation{University of Kansas, Lawrence, Kansas 66045, USA}
\author{T.~K.~Pedlar}
\author{J.~Xavier}
\affiliation{Luther College, Decorah, Iowa 52101, USA}
\author{D.~Cronin-Hennessy}
\author{J.~Hietala}
\author{P.~Zweber}
\affiliation{University of Minnesota, Minneapolis, Minnesota 55455, USA}
\author{S.~Dobbs}
\author{Z.~Metreveli}
\author{K.~K.~Seth}
\author{A.~Tomaradze}
\author{T.~Xiao}
\affiliation{Northwestern University, Evanston, Illinois 60208, USA}
\author{S.~Brisbane}
\author{L.~Martin}
\author{A.~Powell}
\author{P.~Spradlin}
\author{G.~Wilkinson}
\affiliation{University of Oxford, Oxford OX1 3RH, UK}
\author{H.~Mendez}
\affiliation{University of Puerto Rico, Mayaguez, Puerto Rico 00681}
\author{J.~Y.~Ge}
\author{D.~H.~Miller}
\author{I.~P.~J.~Shipsey}
\author{B.~Xin}
\affiliation{Purdue University, West Lafayette, Indiana 47907, USA}
\author{G.~S.~Adams}
\author{D.~Hu}
\author{B.~Moziak}
\author{J.~Napolitano}
\affiliation{Rensselaer Polytechnic Institute, Troy, New York 12180, USA}
\author{K.~M.~Ecklund}
\affiliation{Rice University, Houston, Texas 77005, USA}
\author{J.~Insler}
\author{H.~Muramatsu}
\author{C.~S.~Park}
\author{L.~J.~Pearson}
\author{E.~H.~Thorndike}
\author{F.~Yang}
\affiliation{University of Rochester, Rochester, New York 14627, USA}
\author{S.~Ricciardi}
\affiliation{STFC Rutherford Appleton Laboratory, Chilton, Didcot, Oxfordshire, OX11 0QX, UK}
\author{C.~Thomas}
\affiliation{University of Oxford, Oxford OX1 3RH, UK}
\affiliation{STFC Rutherford Appleton Laboratory, Chilton, Didcot, Oxfordshire, OX11 0QX, UK}
\author{M.~Artuso}
\author{S.~Blusk}
\author{R.~Mountain}
\author{T.~Skwarnicki}
\author{S.~Stone}
\author{J.~C.~Wang}
\author{L.~M.~Zhang}
\affiliation{Syracuse University, Syracuse, New York 13244, USA}
\author{G.~Bonvicini}
\author{D.~Cinabro}
\author{A.~Lincoln}
\author{M.~J.~Smith}
\author{P.~Zhou}
\author{J.~Zhu}
\affiliation{Wayne State University, Detroit, Michigan 48202, USA}
\author{P.~Naik}
\author{J.~Rademacker}
\affiliation{University of Bristol, Bristol BS8 1TL, UK}
\author{D.~M.~Asner}
\altaffiliation[Now at: ]{Pacific Northwest National Laboratory, Richland, WA 99352}
\author{K.~W.~Edwards}
\author{K.~Randrianarivony}
\author{G.~Tatishvili}
\altaffiliation[Now at: ]{Pacific Northwest National Laboratory, Richland, WA 99352}
\affiliation{Carleton University, Ottawa, Ontario, Canada K1S 5B6}
\author{R.~A.~Briere}
\author{H.~Vogel}
\affiliation{Carnegie Mellon University, Pittsburgh, Pennsylvania 15213, USA}
\author{P.~U.~E.~Onyisi}
\author{J.~L.~Rosner}
\affiliation{University of Chicago, Chicago, Illinois 60637, USA}
\author{J.~P.~Alexander}
\author{D.~G.~Cassel}
\author{S.~Das}
\author{R.~Ehrlich}
\author{L.~Fields}
\author{L.~Gibbons}
\author{S.~W.~Gray}
\author{D.~L.~Hartill}
\author{B.~K.~Heltsley}
\author{D.~L.~Kreinick}
\author{V.~E.~Kuznetsov}
\author{J.~R.~Patterson}
\author{D.~Peterson}
\author{D.~Riley}
\author{A.~Ryd}
\author{A.~J.~Sadoff}
\author{X.~Shi}
\author{W.~M.~Sun}
\affiliation{Cornell University, Ithaca, New York 14853, USA}
\author{J.~Yelton}
\affiliation{University of Florida, Gainesville, Florida 32611, USA}
\author{P.~Rubin}
\affiliation{George Mason University, Fairfax, Virginia 22030, USA}
\collaboration{CLEO Collaboration}
\noaffiliation--

\date{June 15, 2011}

\begin{abstract}
We present the results of a Dalitz plot analysis of $D^0 \rightarrow K^0_S\pi^0\pi^0$ using the CLEO-c data set of 818 pb$^{-1}$ of $e^{+}e^{-}$ collisions accumulated at $\sqrt{s} = 3.77~\mathrm{GeV}$.  This corresponds to three million $D^0\overline{D}{}^0$ pairs from which we select 1,259 tagged candidates with a background of $7.5\pm 0.9$ percent.  Several models have been explored, all of which include the $K^*$(892), $K^*_2$(1430), $K^*$(1680), the $f_0$(980), and the $\sigma$(500). We find that the combined $\pi^0\pi^0$ S-wave contribution to our preferred fit is $(28.9\pm 6.3\pm 3.1)$\% of the total decay rate while $D^0 \rightarrow \overline{K}{}^*(892)^0\pi^0$ contributes $(65.6\pm 5.3\pm 2.5)\%$. Using three tag modes and correcting for quantum correlations we measure the $D^0 \rightarrow K^0_S\pi^0\pi^0$ branching fraction to be $(1.059 \pm 0.038 \pm 0.061)\%$.
\end{abstract}

\maketitle

\newcommand{\dzkspppm}{\ensuremath{D^0 \rightarrow K_S^0\pi^+\pi^-}}
\newcommand{\signalDecay}{\ensuremath{D^0 \rightarrow K_S^0\pi^0\pi^0}}
\newcommand{\dzKApz}{\ensuremath{D^0 \rightarrow \overline{K}{}^*(892)^0\pi^0}}
\newcommand{\KAkspz}{\ensuremath{\overline{K}{}^*(892)^0 \rightarrow K_S^0\pi^0}}
\newcommand{\DZkppm}{\ensuremath{\overline{D}{}^0 \rightarrow K^+\pi^-}}
\newcommand{\DZkppmpz}{\ensuremath{\overline{D}{}^0 \rightarrow K^+\pi^-\pi^0}}
\newcommand{\DZkppmpppm}{\ensuremath{\overline{D}{}^0 \rightarrow K^+\pi^-\pi^+\pi^-}}
\newcommand{\DZkppmpppmpz}{\ensuremath{\overline{D}{}^0 \rightarrow K^+\pi^-\pi^+\pi^-\pi^0}}
\newcommand{\pzgg}{\ensuremath{\pi^0 \rightarrow \gamma\gamma}}
\newcommand{\kspppm}{\ensuremath{K_S^0 \rightarrow \pi^+\pi^-}}
\newcommand{\dzksfa}{\ensuremath{D^0 \rightarrow K_S^0f_0(980)}}
\newcommand{\fapzpz}{\ensuremath{f_0(980) \rightarrow \pi^0\pi^0}}
\newcommand{\dzkso}{\ensuremath{D^0 \rightarrow K_S^0\omega}}
\newcommand{\opzg}{\ensuremath{\omega \rightarrow \pz\gamma}}
\newcommand{\kspzpz}{\ensuremath{K_S^0 \rightarrow \pi^0\pi^0}}
\newcommand{\dzksks}{\ensuremath{D^0 \rightarrow K_S^0K_S^0}}
\newcommand{\dppppppm}{\ensuremath{D^+ \rightarrow \pi^+\pi^+\pi^-}}
\newcommand{\dzkppm}{\ensuremath{D^0 \rightarrow K^+\pi^-}}
\newcommand{\dzkppmpz}{\ensuremath{D^0 \rightarrow K^+\pi^-\pi^0}}
\newcommand{\dzkppmpppm}{\ensuremath{D^0 \rightarrow K^+\pi^-\pi^+\pi^-}}
\newcommand{\dzkppmpppmpz}{\ensuremath{D^0 \rightarrow K^+\pi^-\pi^+\pi^-\pi^0}}
\newcommand{\dzKApzsubKAkspz}{\ensuremath{D^0 \rightarrow \overline{K}{}^*(892)(\rightarrow K_S^0\pi^0)\pi^0}}
\newcommand{\dzksfasubfapzpz}{\ensuremath{D^0 \rightarrow K_S^0f_0(980)(\rightarrow \pi^0\pi^0)}}
\newcommand{\dzksosubopzg}{\ensuremath{D^0 \rightarrow K_S^0\omega(\rightarrow \pi^0\gamma)}}
\newcommand{\ks}{\ensuremath{K_S^0}}
\newcommand{\pz}{\ensuremath{\pi^0}}
\newcommand{\epem}{\ensuremath{e^+e^-}}
\newcommand{\psipp}{\ensuremath{\psi(3770)}}
\newcommand{\dz}{\ensuremath{D^0}}
\newcommand{\DZ}{\ensuremath{\overline{D}{}^0}}
\newcommand{\g}{\ensuremath{\gamma}}
\newcommand{\pip}{\ensuremath{\pi^+}}
\newcommand{\pim}{\ensuremath{\pi^-}}
\newcommand{\fa}{\ensuremath{f_0(980)}}
\newcommand{\ka}{\ensuremath{K^*(892)}}
\newcommand{\kb}{\ensuremath{K^*(1680)}}
\newcommand{\kz}{\ensuremath{K_0^*(1430)}}
\newcommand{\ktwo}{\ensuremath{K_2^*(1430)}}
\newcommand{\fb}{\ensuremath{f_0(1370)}}
\newcommand{\fc}{\ensuremath{f_0(1500)}}
\newcommand{\ftwo}{\ensuremath{f_2(1270)}}
\newcommand{\sig}{\ensuremath{\sigma(500)}}
\newcommand{\kp}{\ensuremath{K^+}}
\newcommand{\MeV}{\ensuremath{\mathrm{MeV}}}
\newcommand{\MeVcc}{\ensuremath{\mathrm{MeV/}c^2}}
\newcommand{\GeVGeVcccc}{\ensuremath{\mathrm{GeV^2/}c^4}}
\newcommand{\GeVcc}{\ensuremath{\mathrm{GeV/}c^2}}
\newcommand{\mmkspz}{\ensuremath{m^2(K_S^0\pi^0)}}
\newcommand{\mmpzpz}{\ensuremath{m^2(\pi^0\pi^0)}}
\newcommand{\mbc}{\ensuremath{\mathrm{M(B)}}}
\newcommand{\de}{\ensuremath{\Delta E}}
\newcommand{\Eb}{\ensuremath{E_{\mathrm{beam}}}}
\newcommand{\EbEb}{\ensuremath{E^2_{\mathrm{beam}}}}
\newcommand{\cleocDDlumi}{\ensuremath{\mathrm{818~pb^{-1}}}}
\newcommand{\oldcleocDDlumi}{\ensuremath{\mathrm{281~pb^{-1}}}}

\section{INTRODUCTION}
The substructure of $\dzkspppm$ has been the object of intense recent study due to its relevance to CKM physics \cite{cleoGamma}, but there is currently relatively little information on the substructure of $\signalDecay$. A 1993 CLEO II publication based on 206 events found the branching fraction of the decay chain $\dzKApzsubKAkspz$ to be $(6.7^{+1.8}_{-1.5}) \times 10^{-3}$ and the branching fraction of the non-resonant contribution to be $(4.5 \pm 1.1) \times 10^{-3}$ \cite{Procario}.  The total $\signalDecay$ branching fraction has recently been measured by the CLEO collaboration to be $(8.34 \pm 0.45 \pm 0.42) \times 10^{-3}$ in an analysis that used $\oldcleocDDlumi$ of data at $\sqrt s=3.77$ GeV \cite{cleoDD}.

The pursuit of a more comprehensive study of the substructure of $\signalDecay$ is motivated in part by the fact that this mode should provide a cleaner way to investigate the $\pi\pi$ S-wave substructure observed in $\dzkspppm$ \cite{Muramatsu,babar,belle}.  Analyses performed by both BaBar and Belle use eight $\pi\pi$ resonances, four of which are spin-zero (including two distinct $\sigma$ resonances), to fit the $\dzkspppm$ Dalitz plot, but the $\pi\pi$ S-wave components in these analyses are masked by the very sizeable P-wave contributions of the $\rho$ resonance.  By studying $\ks\pz\pz$ we eliminate final states involving $\rho$ mesons and expect any S-wave structure present in the decay to be more prominent.

\section{EXPERIMENTAL TECHNIQUE}
\begin{figure}
\includegraphics*[width=3.00in]{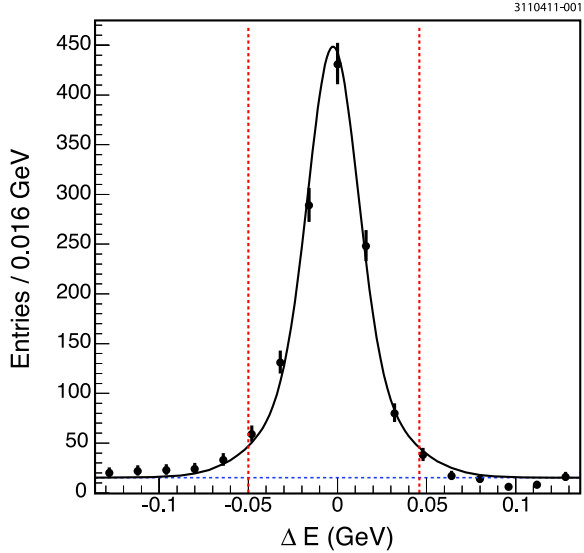}
\caption{Signal-side $\de$ distribution.  The solid line shows the results of fitting with a double Gaussian to represent the signal plus a constant background (horizontal dashed line). The vertical dashed lines indicate the signal region.}
\label{fig:DE}
\end{figure}

We have analyzed $\cleocDDlumi$ of $\epem$ collisions produced with the Cornell Electron Storage Ring at the center-of-mass energy of the $\psipp$ resonance, resulting in about three million $\dz\DZ$ pairs produced in the CLEO-c detector.  We consider candidates where the $\dz$ is reconstructed as $\ks\pz\pz$, and the $\DZ$ is reconstructed in the following four tag modes: $\DZkppm$, $\DZkppmpz$, $\DZkppmpppm$, and $\DZkppmpppmpz$ (charge congugation is implied throughout this paper unless explicitly stated).  Reconstructing any of these modes, whether it be signal or tag, involves placing requirements on the beam-constrained mass $m(D)$ and the energy difference $\de$, defined as
\begin{eqnarray}
  \label{eqn:MBC_def}
  m(D)&=&\sqrt{\EbEb/c^4-p_D^2/c^2}\\
  \label{eqn:DE_def}
  \de&=&E_D-\Eb
\end{eqnarray}
where $\Eb$ is one-half of the center-of-mass energy of the colliding beams, and $E_D$ ($p_D$) is the total reconstructed energy (momentum) of the candidate.  Both of these quantities rely on the fact that $E_D$ would be the same as $\Eb$ if the event were perfectly reconstructed, and are required to be within 2.5 (3.5) standard deviations of their nominal values for $m(D)$ ($\de$).  Figure \ref{fig:DE} shows the $\de$ distributions for tagged $\signalDecay$ candidates that pass all other selection criteria.

We reconstruct neutral pions using photons with $E_\gamma > 30~\MeV$ and $m_{\gamma\gamma}$ within 3 standard deviations (15 $\MeVcc$) of the nominal $\pz$ mass~\cite{PDG}.  We reconstruct $\kspppm$ with $m_{\pip\pim}$ within 2.5 standard deviations (6 $\MeVcc$) of the nominal $\ks$ mass~\cite{PDG} and with the $\pip\pim$ vertex displaced from the interaction point by at least twice the error on their separation.  When there are multiple $\signalDecay$ candidates in a single event we select the one with the highest joint probability that both neutral pions are correctly identified.

\begin{figure}
\includegraphics*[width=3.00in]{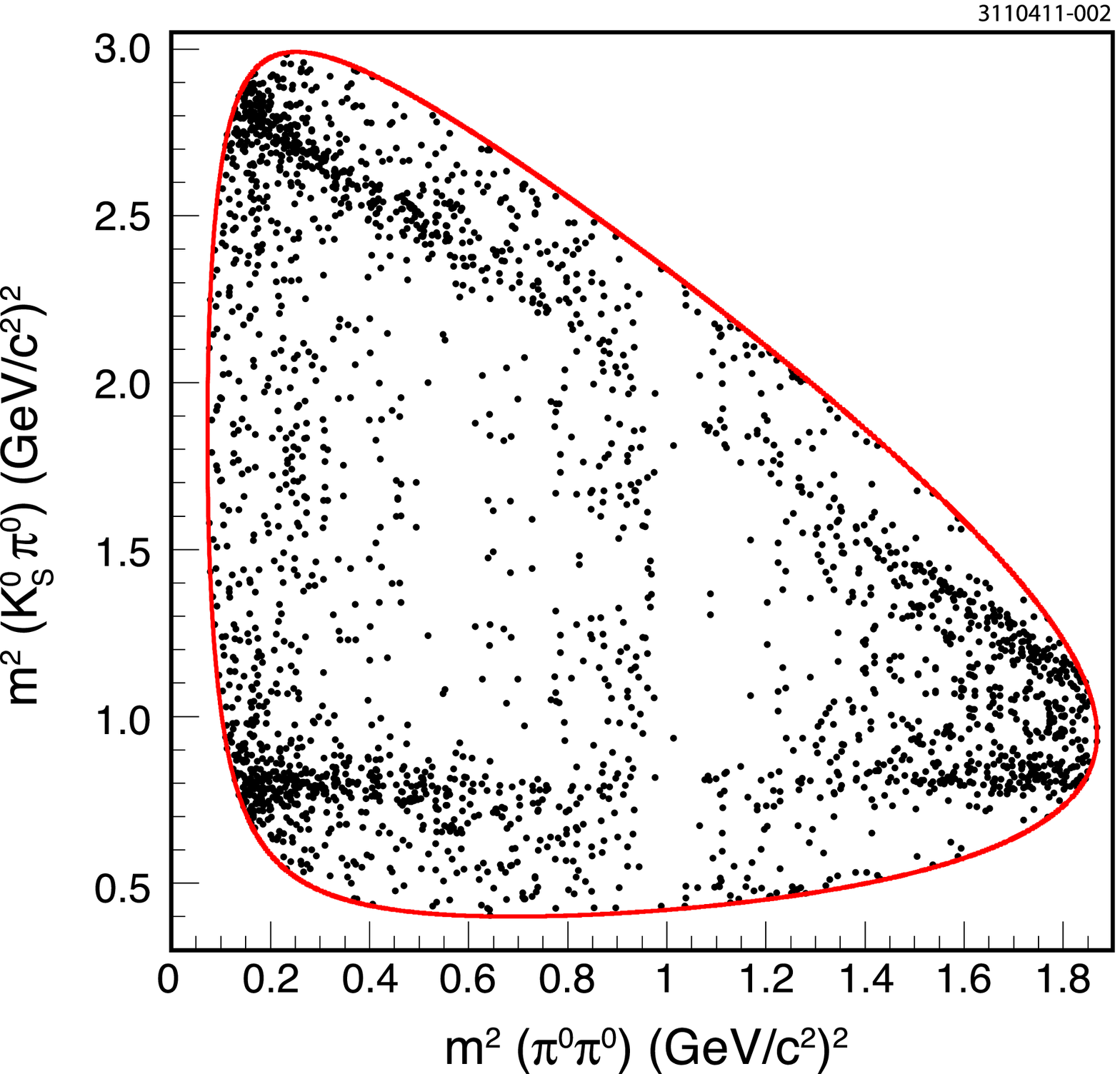}
\caption{$\signalDecay$ Dalitz plot showing two entries per candidate, one for each possible $\ks\pz$ combination.}
\label{fig:DP2zy}
\end{figure}

After all selection requirements are imposed on both signal and tag-side candidates we obtain a combined sample of 1,259 tagged $\signalDecay$ events.  A Dalitz plot of these candidates is shown in Fig. \ref{fig:DP2zy}.  Bose symmetry requires that the decay dynamics remain invariant when the two $\pz$'s are swapped, hence the Dalitz plot contains two entries per event, one for each possible $\ks\pz$ combination.  The horizontal band evident at around $\mmkspz = 0.8~\GeVGeVcccc$ is due to $D^0\rightarrow\overline{K}{}^*(892)\pi^0$ ($\overline{K}{}^*(892)\rightarrow K^0_S\pi^0$), and the most striking vertical feature is an empty band at around $\mmpzpz = 1~\GeVGeVcccc$ where the $K\overline{K}$ threshold opens, indicating the presence of destructive interference between the $\fa$ and some broad underlying $\pi\pi$ S-wave structure.  A study of events both inside and outside the signal region indicated in Fig. \ref{fig:DE} shows that  $(7.5 \pm 0.9)\%$ of the events in this Dalitz plot are background.

We determine the efficiency of our analysis as a function of Dalitz plot coordinate by using a GEANT-based Monte Carlo package to generate 1,000,000 simulated $\dz\DZ$ pairs where one $D$ is forced to decay to a flavor-tagging mode in proportion to its branching fraction while the other $D$ decays to $\ks\pz\pz$ uniformly across the Dalitz plot.  We pass these events through the same analysis code and selection requirements as the data, and fit the Dalitz plot of the remaining 63,391 events to a cubic polynomial which has been explicitly symmetrized in the two $\mmkspz$ variables.  We find that the efficiency is well modeled by this simple function and is uniform across the Dalitz plot.

To study backgrounds we use a Monte Carlo-generated sample of $\dz\DZ$ events corresponding to twenty times the actual integrated luminosity in which the $D$'s decay to all experimentally measured final states with appropriate branching fractions, removing all events that contain an actual $\signalDecay$ decay and subjecting the remaining events to the same analysis requirements as the real data.  The remaining background events vary smoothly across the Dalitz plot, preferring the corners of phase space where a $\pz$ is produced at rest and is more easily faked.  We find the background shape is well modeled by a symmetrized third-order polynomial combined with a small Breit-Wigner component to account for $\omega$ decays from the background mode $\dzksosubopzg$ where the $\omega$ decay combines with an unrelated calorimeter shower to yield a second $\pz$ candidate.

\section{DALITZ PLOT ANALYSIS}
The fitting method used to explore the efficiency and background shapes, as well as the structure of the $\signalDecay$ signal, uses the same unbinned maximum likelihood technique described in Ref. \cite{Tim} which minimizes the sum over N events:
\begin{equation}\label{eqn:LogL}
  \mathcal{L}=-2\sum_{n=1}^{N} \log {\cal P}(x_n,y_n)
\end{equation}
where $x$ and $y$ are the two $\mmkspz$ Dalitz variables and ${\cal P}(x,y)$ is the probability density function (p.d.f.) which depends on the event sample being fit:
\begin{equation}\label{eqn:PDF}
  {\cal P}(x,y) = \left\{
  \begin{array}{ll}
    {\mathcal N}_{\varepsilon}\varepsilon(x,y) & {\rm for~efficiency;}\\
    {\mathcal N}_B B(x,y) & {\rm for~background;}\\
    f_{\rm{sig}} {\mathcal N}_S |{\mathcal M}(x,y)|^2 \varepsilon(x,y) + (1-f_{\rm{sig}}) {\mathcal N}_B B(x,y) & {\rm for~signal.}
  \end{array}\right.
\end{equation}
The signal p.d.f. is proportional to the matrix element squared, $|{\mathcal M}(x,y)|^2$, corrected by the measured efficiency $\varepsilon(x,y)$ and the signal fraction, $f_{\rm{sig}} = (0.924\pm 0.009)$, which is simply the complement of the background fraction discussed above, and is fixed in the fit.  The efficiency, signal, and background contributions are normalized separately:
\begin{eqnarray}
  \label{eqn:eff_normalization}
  {1\over{{\mathcal N}_{\varepsilon}}} &=& \int \varepsilon(x,y) dx dy,\\
  \label{eqn:signal_normalization}
  {1\over{{\mathcal N}_S}} &=& \int |{\mathcal M}(x,y)|^2 \varepsilon(x,y) dx dy,\\
  \label{eqn:bkg_normalization}
  {1\over{{\mathcal N}_B}} &=& \int B(x,y) dx dy,
\end{eqnarray}
providing the overall normalization $\int {\cal P}(x,y) dx dy = 1$.

Several models for fitting the data have been explored, and three of these are described here.  All include the $\ka$, $\ktwo$, and $\kb$ as intermediate $\ks\pz$ resonances and an $\ftwo$ intermediate $\pz\pz$ resonance.  To model the S-wave contribution to $\pz\pz$ we add an $\fa$, described with a Flatt\'{e} parameterization \cite{flatte}, to a $\pi\pi$ complex pole around the mass of the $\sig$, and we include one of the following three variations: $\fb$ (Model 1), $\fc$ (Model 2), and both $\fb$ and $\fc$ (Model 3).  Each includes a non-interfering $\kspzpz$ contribution to account for $\dzksks$.  A fourth model, in which the broadest S-wave features were replaced by a simple non-resonant component, was rejected since the fit fraction of the non-resonant component exceeded $80$\%. Table \ref{tbl:resonance_parameters} summarizes the parameters used for the intermediate resonances and the functions used to represent them.

\begin{table}
\caption{Details of the resonances included in the Dalitz plot fit.}
\vspace*{0.5ex}
\label{tbl:resonance_parameters}
\begin{tabular}{cccc}
  \hline \hline
  Resonance
    & Model
    & Mass ($\MeVcc$)
    & Width ($\MeVcc$)\\
  \hline
  $\pi\pi$ pole
    & Complex Pole
    & $470 - i220$
    &\\
  $\ka$
    & Breit-Wigner
    & 896
    & 50.3\\
  $\fa$
    & Flatt\'{e}\footnote{$g_{\pi\pi} = 406~\MeVcc$, $g_{KK}/g_{\pi\pi} = 2$}
    & 965
    &\\
  $\ftwo$
    & Breit-Wigner
    & 1275.1
    & 185.0\\
  $\fb$
    & Breit-Wigner
    & 1350
    & 265\\
  $\ktwo$
    & Breit-Wigner
    & 1432.4
    & 109\\
  $\fc$
    & Breit-Wigner
    & 1505
    & 109\\
  $\kb$
    & Breit-Wigner
    & 1717
    & 322\\
  \hline \hline
\end{tabular}
\end{table}

Table \ref{tbl:dpfit_data_ind} summarizes the Dalitz plot fit results for the three models described above.  In all cases we choose the amplitude (phase) of the $\ka$ to be 1 (0) respectively, which effectively defines the complex coordinate system for the other resonances included in the fit.  The $\chi^2$ values shown are calculated post-fit by dividing the Dalitz plot into discrete bins and comparing the data with the average p.d.f. in each bin.  The fit fraction (FF) for each component is calculated by integrating its contribution across the Dalitz plot and dividing by the integral of the coherent sum of all components:
\begin{equation}\label{eqn:FF}
FF_i = {{\int_{\cal DP} |a_i|^2}\over {\int_{\cal DP} |\sum_j a_je^{i\phi_j}|^2}},
\end{equation}
where $a_j$ ($\phi_j$) are the amplitude (phase) of the $j$-th component. The total fit fraction for a given model is the sum of the fit fractions for the individual components of that model, and a total fit fraction greater than unity indicates the presence of destructive interference on the Dalitz plot.

\begin{figure}
\begin{minipage}[t]{3.00in}
\begin{center}
\includegraphics*[width=3.00in]{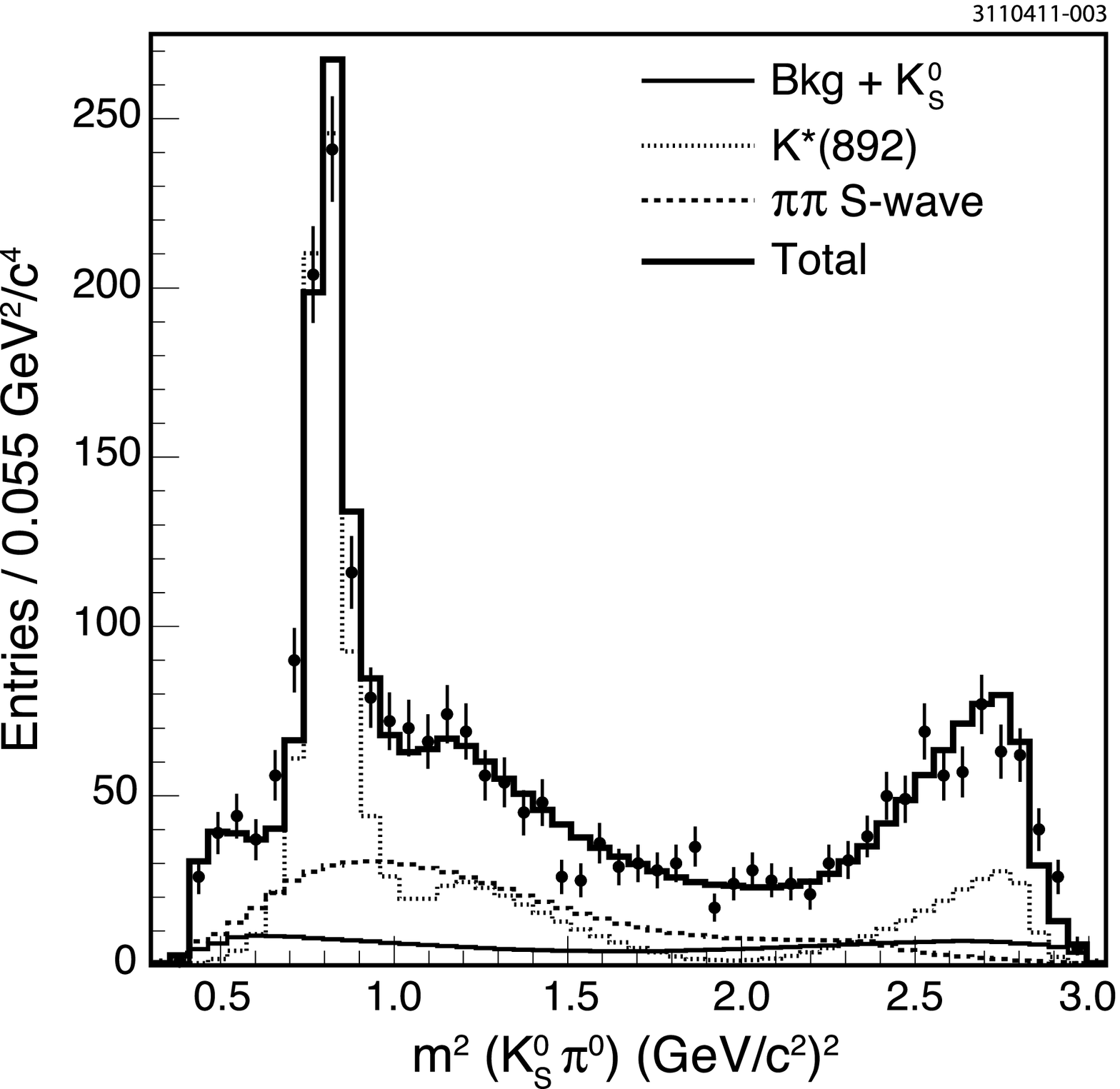}
\end{center}
\caption{$m^2(K\pi)$ projections for Model 1.  The data are represented  by the points with error bars and the total fit by the thick solid line. The other fit components are discussed in the text.}
\label{fig:dpfit_data_m1_XY}
\end{minipage}
\hfill
\begin{minipage}[t]{3.00in}
\begin{center}
\includegraphics*[width=3.00in]{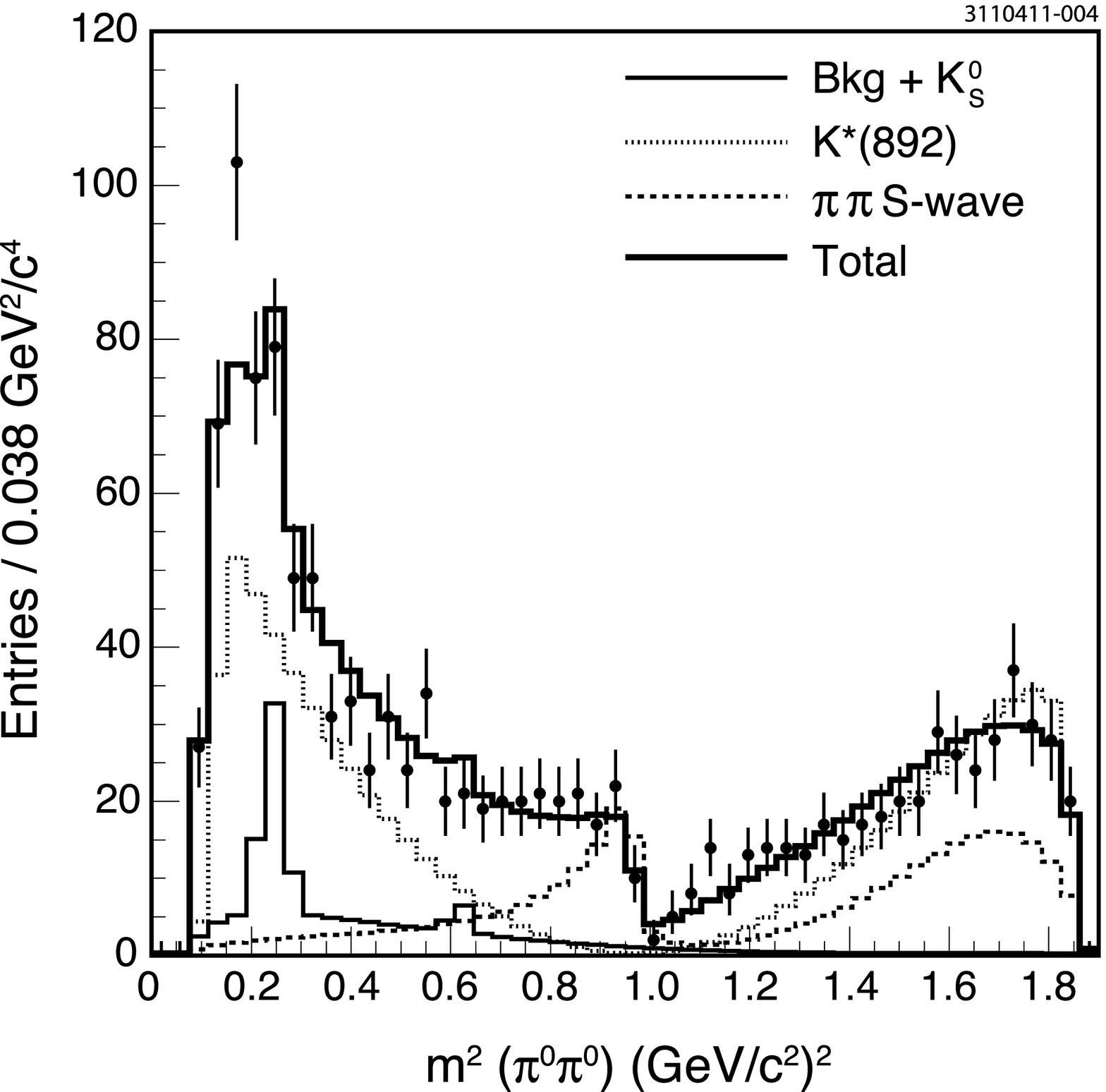}
\end{center}
\caption{$m^2(\pi\pi)$ projection for Model 1.  The data are represented  by the points with error bars and the total fit by the thick solid line. The other fit components are discussed in the text.}
\label{fig:dpfit_data_m1_Z}
\end{minipage}
\end{figure}

While the quality of the fit projections of the three models are comparable, we choose Model 1 as our preferred fit since it has both a reasonable $\chi^2$ (unlike Model 2) and a total fit fraction close to unity (unlike Model 3). Figures \ref{fig:dpfit_data_m1_XY} and \ref{fig:dpfit_data_m1_Z} show the projections of both data and p.d.f. for Model 1. In both figures the data are represented  by the points with error bars, the total fit by the thick solid line, the background contribution by the thin solid line, the $\pi\pi$ S-wave contribution by the dashed line, and the $K^*(892)$ contribution by the broken solid line.
\begin{table}
\caption{Fit fraction (\%), amplitude, and phase ($^\circ$) fit results for Models 1 through 3.  The errors are statistical only.}
\vspace*{0.5ex}
\label{tbl:dpfit_data_ind}
\begin{tabular}{ccr@{$\pm$}lr@{$\pm$}lr@{$\pm$}l}
  \hline \hline
  \multicolumn{2}{c}{}
    & \multicolumn{2}{c}{Model 1}
    & \multicolumn{2}{c}{Model 2}
    & \multicolumn{2}{c}{Model 3}\\
  \hline
  \multicolumn{2}{c}{$\chi^{2}/N_{dof}$}
    & \multicolumn{2}{c}{$19.9/7$}
    & \multicolumn{2}{c}{$44.7/7$}
    & \multicolumn{2}{c}{$14.9/5$}\\
  \multicolumn{2}{c}{Total FF}
    & 122 & 8
    & 120 & 11
    & 252 & 33\\
  \hline \hline
    & FF
      & 2.7  & 1.4
      & 3.7  & 1.8
      & 4.5  & 2.2\\
  $\pi\pi$ pole
    & $a$
      & 0.67 & 0.16
      & 0.91 & 0.20
      & 0.99 & 0.23\\
    & $\phi$
      & 140  & 17
      & 119  & 22
      & 39   & 17\\
  \hline
    & FF
      & 10.5 & 2.1
      & 12.1 & 2.4
      & 18.4 & 4.3\\
  $\fa$
    & $a$
      & 1.71 & 0.17
      & 2.13 & 0.20
      & 2.59 & 0.24\\
    & $\phi$
      & 35.2 & 9.9
      & 65   & 11
      & 44.8 & 7.9\\
  \hline
    & FF
      & 25.7 & 5.1
      & \multicolumn{2}{c}{(n/a)}
      & 81   & 24\\
  $\fb$
    & $a$
      & 5.72 & 0.58
      & \multicolumn{2}{c}{(n/a)}
      & 11.6 & 1.5\\
    & $\phi$
      & 340.3 & 6.6
      & \multicolumn{2}{c}{(n/a)}
      & 15.8  & 8.6\\
  \hline
    & FF
      & \multicolumn{2}{c}{(n/a)}
      & 22.3 & 6.0
      & 73   & 20\\
  $\fc$
    & $a$
      & \multicolumn{2}{c}{(n/a)}
      & 11.7 & 1.5
      & 20.9 & 4.0\\
    & $\phi$
      & \multicolumn{2}{c}{(n/a)}
      & 16    & 12
      & 281.4 & 8.0\\
  \hline
    & FF
      & 2.48  & 0.91
      & 12.9  & 3.3
      & 6.8   & 2.5\\
  $\ftwo$
    & $a$
      & 1.57 & 0.28
      & 4.16 & 0.54
      & 2.98 & 0.53\\
    & $\phi$
      & 282   & 18
      & 2.2   & 6.5
      & 340.9 & 8.9\\
  \hline \hline
    & FF
      & 65.6 & 5.3
      & 48.6 & 5.9
      & 50.2 & 7.8\\
  $\ka$
    & $a$
      & \multicolumn{2}{c}{1 (fixed)}
      & \multicolumn{2}{c}{1 (fixed)}
      & \multicolumn{2}{c}{1 (fixed)}\\
    & $\phi$
      & \multicolumn{2}{c}{0 (fixed)}
      & \multicolumn{2}{c}{0 (fixed)}
      & \multicolumn{2}{c}{0 (fixed)}\\
  \hline
    & FF
      & 0.49 & 0.45
      & 1.9  & 1.2
      & 1.45 & 0.82\\
  $\ktwo$
    & $a$
      & 0.43 & 0.18
      & 0.98 & 0.29
      & 0.85 & 0.24\\
    & $\phi$
      & 141 & 28
      & 191 & 16
      & 159 & 15\\
  \hline
    & FF
      & 11.2  & 2.7
      & 15.2  & 4.3
      & 13.5  & 3.5\\
  $\kb$
    & $a$
      & 5.65 & 0.77
      & 7.6  & 1.2
      & 7.07 & 0.82\\
    & $\phi$
      & 55   & 11
      & 45   & 12
      & 18.7 & 8.9\\
  \hline \hline
    & FF
      & 3.46 & 0.92
      & 3.30 & 0.96
      & 2.48 & 0.95\\
  $\ks$
    & $a$
      & 0.281 & 0.037
      & 0.318 & 0.044
      & 0.272 & 0.050\\
    & $\phi$
      & \multicolumn{6}{c}{contributes incoherently} \\
  \hline \hline
\end{tabular}
\end{table}

The results in Table~\ref{tbl:dpfit_data_ind} show a large model dependence of the fit fractions and phases of both the $\fa$ and the $\pi\pi$ pole while the overall fit projections remain largely unchanged.  We interpret this as an indication that, while the overall shape and phase evolution of the $\pi\pi$ S-wave component of our fits are reasonable, the details of the underlying physics model used to describe this S-wave component are not well determined.  To test the hypothesis that the combined structure of the S-wave components is meaningful while the specific details of these models are not, we combine all of the individual spin-zero $\pz\pz$ components into a single $\pi\pi$ ``S-wave object'' and calculate the fit fraction of this object as a whole:
\begin{equation}\label{eqn:FF_Swave}
FF_{S-wave} = {{\int_{\cal DP} |\sum_{S-wave} a_ke^{i\phi_k}|^2}\over {\int_{\cal DP} |\sum_{all} a_je^{i\phi_j}|^2}},
\end{equation}
The results summarized in Table~\ref{tbl:dpfit_data} show that the $\pi\pi$ S-wave fit fraction is now consistent among models, as are the total fit fractions which are now near 100\%.
\begin{table}
\caption{Dalitz plot fit fractions (\%) with a coherent $\pi\pi$ S-wave object.}
\vspace*{0.5ex}
\label{tbl:dpfit_data}
\begin{tabular}{cr@{$\pm$}lr@{$\pm$}lr@{$\pm$}l}
  \hline \hline
    & \multicolumn{2}{c}{Model 1}
    & \multicolumn{2}{c}{Model 2}
    & \multicolumn{2}{c}{Model 3}\\
  \hline
  $\pi\pi$ S-wave
    & 28.9 & 6.3
    & 19.5 & 5.7
    & 30   & 12\\
  $\ka$
    & 65.6 & 5.3
    & 48.6 & 5.9
    & 50.2 & 7.8\\
  $\ktwo$
    & 0.49 & 0.45
    & 1.9  & 1.2
    & 1.45 & 0.82\\
  $\kb$
    & 11.2 & 2.7
    & 15.2 & 4.3
    & 13.5 & 3.5\\
  $\ftwo$
    & 2.48 & 0.91
    & 12.9 & 3.3
    & 6.8  & 2.5\\
  $\ks$
    & 3.46 & 0.92
    & 3.30 & 0.96
    & 2.48 & 0.95\\
  \hline
  Total
    & 112.1 & 8.8
    & 101   & 10
    & 105   & 15\\
  \hline \hline
\end{tabular}
\end{table}

The fact that the underlying physics is not well represented by a simple combination of resonances is not surprising given the results of other analysis efforts where ``$\sigma$'' like objects have turned out to be the result of dynamical effects that require a more sophisticated modeling approach \cite{focus}.  In spite of this shortcoming, our simple model does capture the striking overall feature that there is significant destructive interference at $\mmpzpz$ around 1 $\GeVGeVcccc$, and that the overall $\pz\pz$ S-wave component accounts for about a quarter of all $\signalDecay$ decays. These results are consistent with the studies of $\dzkspppm$ from BaBar and Belle in which the total $\pi\pi$ S-wave plus non-resonant fit fractions were found to be $27.2$\% and $26.1$\% respectively \cite{babar,belle}.

To test the stability of our results as selection requirements are varied, we tightened (loosened) these requirements by half a standard deviation resulting in 1,099 (1,361) candidates in the Dalitz plot.  In each case the signal fraction $f_{\rm{sig}}$, efficiency shape $\varepsilon(x,y)$, and background shape $B(x,y)$ were remeasured and fixed in the fit to the Dalitz plot, which used Model 1 as described above.  Sensitivity to uncertainties in identifying and reconstructing $\pz$s was studied by tightening the nominal pull mass requirement to $\pm 2.5$ standard deviations, resulting in 1,150 events in the Dalitz plot, and by eliminating all $\pz$s having one or more photons in the endcap calorimeter ($\cos\theta > 0.81$), resulting in 1,023 events in the Dalitz plot. In the same way, sensitivity to the signal fraction was studied by varying $f_{\rm{sig}}$ by $\pm 3$ standard deviations from its nominal value; sensitivity to the shape of the efficiency function was studied by replacing the fitted efficiency shape parameters with those corresponding to a uniform efficiency $\varepsilon(x,y) = 1$; and sensitivity to uncertainty in the shape of the background was studied by replacing the nominal background shape, determined using Monte Carlo-generated data, with a background shape determined by analyzing single tagged $\signalDecay$ data selected from a sideband region in the $\de$ vs. $m(D)$ plane.

In all cases described above, the parameters derived from fitting the resulting Dalitz plot with Model 1 were within 1 standard deviation of the parameters found in the preferred fit. For each parameter we choose the maximum variation from the preferred fit value to represent the systematic uncertainty in that parameter.  The resulting systematic errors for the fit fractions from Dalitz plot Model 1 are given in \ref{tbl:dalitz_result}.

\begin{table}
\caption{Dalitz plot fit results for Model 1. The first error is statistical and the second represents variations on analysis requirements.}
\vspace*{0.5ex}
\label{tbl:dalitz_result}
\begin{tabular}{cr@{$~\pm~$}r@{.}l@{$~\pm~$}r@{.}l}
  \hline \hline
  Component
    & \multicolumn{5}{c}{Fit Fraction (\%)}\\
  \hline
  $\pi\pi$ S-wave
    & 28.9
    &  6&3
    &  3&1\\
  $\ka$
    & 65.6
    &  5&3
    &  2&5\\
  $\ktwo$
    & 0.49
    & 0&45
    & 0&23\\
  $\kb$
    & 11.2
    &  2&7
    &  2&5 \\
  $\ftwo$
    & 2.48
    & 0&91
    & 0&78\\
  $\ks$
    & 3.46
    & 0&92
    & 0&66\\
  \hline \hline
\end{tabular}
\end{table}

\section{BRANCHING FRACTION ANALYSIS}
The extraction of the $\signalDecay$ branching fraction is a straightforward extension of the Dalitz analysis described above; we need only to quantify the yields and efficiencies for observing the signal $\signalDecay$ events as a function of the tag mode.  More specifically, to extract the $\signalDecay$ branching fraction using double-tagged events where one $D$ decays to $\ks\pz\pz$ and the other $D$ decays to tag mode $\tau$, we need to evaluate
\begin{equation}\label{eqn:br1}
  \mathcal{B}(\signalDecay)_\tau = \frac{N^\tau_{DT} / (\varepsilon^\tau_{DT} \alpha_\tau)} {2N_{\dz\DZ}\mathcal{B}_\tau},
\end{equation}
where $\varepsilon^\tau_{DT}$ is the efficiency for the detector and analysis to select an event where the $\dz$ decays to $\ks\pz\pz$ and the $\DZ$ decays to tag mode $\tau$, $N^\tau_{DT}$ is the measured yield of these events, $N_{\dz\DZ}$ is the number of $\dz\DZ$ pairs produced, $\mathcal{B}_\tau$ is the branching fraction of the tag mode, $\alpha_\tau$ is a small correction to account for the quantum correlation between the two sides of the event, and the factor of 2 accounts for the fact that we include charge-conjugate decays.

The yield of $\signalDecay$ signal events for each tag mode, $N^\tau_{DT}$, is determined by fitting the $\de$ distribution for the tagged events to a double Gaussian signal plus a flat background as illustrated in Fig. \ref{fig:DE}.  The yields for all tag modes are summarized in Table \ref{tbl:numbers}.

The efficiency for detecting a $\dz$ decaying to $\ks\pz\pz$ when the $\DZ$ decayed to tag mode $\tau$ was determined by generating a large sample of simulated $\dz\DZ$ events for each case and measuring the fraction of these events that pass the event selection requirements.  The numbers are summarized in Table \ref{tbl:numbers}, and include an additional factor of $\varepsilon^{corr}_{\pz} = 0.94 \pm 0.02$ for each $\pz$ to correct for a known systematic difference between data and Monte Carlo.

The factor $2N_{\dz\DZ}\mathcal{B}_\tau$ in the denominator of Eq.~\ref{eqn:br1} is number of the $D$'s that decayed to tag mode $\tau$ in the CLEO-c detector.  This is determined from the efficiency-corrected single-tag yields obtained by a recently published CLEO-c analysis that used the same data sample and tag modes to study semileptonic charm decays \cite{tag_yields}.  The numbers are summarized in Table \ref{tbl:numbers}.

The correction factor $\alpha_\tau$ is needed because the signal-side $D$ decays to a $CP$-even eigenstate and the tag-side $D$ decays to a flavor eigenstate of mixed $CP$, allowing interference between favored and suppressed decay paths to modify the observed decay rate at the five percent level.  The formalism is discussed in Refs. \cite{cleoDD} and \cite{qcTwo} where $\alpha_\tau$ is defined as
\begin{equation}
  \alpha_\tau = 1 + 2r\cos\delta_\tau + R_{WS}^\tau + y \equiv 1 + R_{WS}^\tau + \Delta_{QC}^\tau
\end{equation}
and $\delta_\tau$ is the strong phase difference between the amplitudes of $\dz$ and $\DZ$ decaying to tag mode $\tau$, $R_{WS}^\tau$ is the relative rate of $\DZ$ and $\dz$ decaying to tag mode $\tau$, and $y$ is the standard $\dz\DZ$ mixing parameter while the values of $R_{WS}^\tau$ and $\Delta_{QC}^\tau$ for the $\dzkppm$ tag mode are obtained from the Particle Data Group \cite{PDG}, and the values for the $\dzkppmpz$ and $\dzkppmpppm$ tag modes are obtained from Ref.~\cite{qcTwo}.  Since $R_{WS}^\tau$ and $\Delta_{QC}^\tau$ are not known for $\dzkppmpppmpz$, this tag mode is not used to extract the $\signalDecay$ branching fraction.  The overall correction factor $\alpha_\tau$ for each tag mode is reported in Table~\ref{tbl:numbers}. We calculate the $\signalDecay$ branching fraction for each tag mode and show the results in Table~\ref{tbl:BR}.

 To study the systematic uncertainty associated with these measurements we varied the way both the signal and background are parameterized when extracting signal yields, in all cases performing the same procedure on both data and Monte Carlo.  Other analysis variations were also explored, including the tightening and loosening of reconstruction requirements as described in the previous section.  These variations were combined in quadrature with the uncertainties in $\alpha$, $\varepsilon_{\pz}^{corr}$, $N_{ST}/\varepsilon_{ST}$, tracking, $\ks$ efficiency, and the statistical uncertainty in measuring $\varepsilon_{DT}$, to give the total systematic uncertainty for each tag mode.  Table \ref{tbl:BR} lists the results from our branching ratio calculations and gives our final result by performing a weighted average across the three modes.
\begin{table}
\caption{The yields, efficiency, and quantum correction for each tag mode.}
\vspace*{0.5ex}
\label{tbl:numbers}
\begin{tabular}{cr@{$\pm$}lr@{$\pm$}lr@{$\pm$}lr@{$\pm$}lr@{$\pm$}l}
  \hline \hline
  Tag Mode
    & \multicolumn{2}{c}{$N^\tau_{DT}$}
    & \multicolumn{2}{c}{$\varepsilon^\tau_{DT}$ (\%)}
    & \multicolumn{2}{c}{$2N_{\dz\DZ}\mathcal{B}_\tau$}
    & \multicolumn{2}{c}{$\alpha_\tau$}\\
  \hline
  $\kp\pim$
    & 247    & 17
    & 9.29   & 0.41
    & 229050 &  600
    & 1.130  & 0.041\\
  $\kp\pim\pz$
    & 500    & 25
    & 5.30   & 0.26
    & 809700 & 1700
    & 1.099  & 0.027\\
  $\kp\pim\pip\pim$
    & 358    & 28
    & 6.04   & 0.30
    & 449500 & 1100
    & 1.080  & 0.028\\
  \hline \hline
\end{tabular}
\end{table}
\begin{table}
\caption{Summary of $\mathcal{B}(\signalDecay)$ measurements.}
\vspace*{0.5ex}
\label{tbl:BR}
\begin{tabular}{cc}
  \hline \hline
  Tag Mode
    & $\mathcal{B}(\signalDecay)$ (\%)\\
  \hline
  $\kp\pz$
    & $1.030 \pm 0.069 \pm 0.094$\\
  $\kp\pim\pz$
    & $1.061 \pm 0.054 \pm 0.110$\\
  $\kp\pim\pip\pim$
    & $1.099 \pm 0.084 \pm 0.115$\\
  \hline
  Average
    & $1.059 \pm 0.038 \pm 0.061$ \\
  \hline \hline
\end{tabular}
\end{table}

\section{CONCLUSION}
We have performed a Dalitz plot analysis of the decay mode $\signalDecay$ using the full CLEO-c data set and have measured the total branching fraction of this mode to be $(1.059 \pm 0.038 \pm 0.061)\%$.  We find that the combined $\pi^0\pi^0$ S-wave contribution to our preferred fit is $(28.9\pm 6.3\pm 3.1)$\% of the total decay rate while $D^0 \rightarrow \overline{K}{}^*(892)^0\pi^0$ contributes $(65.6\pm 5.3\pm 2.5)\%$.

\begin{acknowledgments}
We gratefully acknowledge the effort of the CESR staff
in providing us with excellent luminosity and running conditions.
D.~Cronin-Hennessy thanks the A.P.~Sloan Foundation.
This work was supported by
the National Science Foundation,
the U.S. Department of Energy,
the Natural Sciences and Engineering Research Council of Canada, and
the U.K. Science and Technology Facilities Council.
\end{acknowledgments}


\begin{thebibliography}{99}

\bibitem{cleoGamma}
See, for example, R.~Briere {\it et~al.} {(CLEO Collaboration)}, {Phys. Rev.} {D {\bf80}}, {032002} ({2009}),
and  J.~Libby {\it et~al.} {(CLEO Collaboration)}, {Phys. Rev.} {D {\bf82}}, {112006} ({2010}).

\bibitem{Procario}
M.~Procario {\it et~al.} {(CLEO Collaboration)}, {Phys. Rev.} {D {\bf48}}, {4007} ({1993}).

\bibitem{cleoDD}
D.~Asner {\it et~al.} {(CLEO Collaboration)}, {Phys. Rev.} {D {\bf78}}, {012001} ({2008}).

\bibitem{Muramatsu}
  H.~Muramatsu {\it et al.}  {(CLEO Collaboration)}, {Phys.Rev.Lett.} {\bf 89}, 251802 (2002)
  {and Erratum}  {\bf 90}, {059901} {(2003)}

\bibitem{babar}
B.~Aubert {\it et~al.} {(Babar Collaboration)}, {Phys. Rev.} {Lett. {\bf95}}, 121802 ({2005}).

\bibitem{belle}
A.~Poluektov {\it et~al.} {(Belle Collaboration)}, {Phys. Rev.} {D {\bf73}}, {112009} ({2006}).

\bibitem{PDG}
C.~Amsler {\it et~al.}, {Phys. Lett.} {B {\bf667}}, {1} {(2008) and 2009 partial update for the 2010 edition}.

\bibitem{Tim}
S.~Kopp {\it et~al.} {(CLEO Collaboration)}, {Phys. Rev.} {D {\bf63}}, {092001} ({2001}).

\bibitem{flatte}
S.M.~Flatt\'{e}, {Phys. Lett.} {B {\bf63}}, 224 ({1976}).

\bibitem{Dubrovin_threePi}
G. ~Bonvicini {\it et~al.} {(CLEO Collaboration)}, {Phys. Rev.} {D {\bf76}}, 012001 {(2007)}.

\bibitem{focus}
A.~Link {\it et al.} {(FOCUS Collaboration)}, {Phys. Lett.} {B {\bf653}}, 1 ({2007}).

\bibitem{tag_yields}
D.~Besson {\it et~al.} {(CLEO Collaboration)}, {Phys. Rev.} {D {\bf80}}, {032005} ({2009}).

\bibitem{qcTwo}
N.~Lowrey {\it et~al.} {(CLEO Collaboration)}, {Phys. Rev.} {D {\bf80}}, {031105(R)} ({2009}).

\end{thebibliography}
\end{document}